\documentclass[aps,prl,twocolumn,showpacs,superscriptaddress]{revtex4-1}
\usepackage{graphicx}
\usepackage{amsmath}
\usepackage{amssymb}
\usepackage{colordvi}
\usepackage{mathrsfs}
\usepackage{bm}
\usepackage{verbatim}
\usepackage{dcolumn}
\usepackage{bm}
\usepackage{epsfig}
\usepackage{subfigure}
\usepackage{dsfont}

\usepackage[colorlinks,linkcolor=blue,anchorcolor=blue,citecolor=blue]{hyperref}

\begin{document}
\title{Van der Waals Heterostructure Pt$_{2}$HgSe$_{3}$/CrI$_3$ for Topological Valleytronics}
\author{Zheng Liu$^\dag$}
\affiliation{ICQD, Hefei National Laboratory for Physical Sciences at Microscale, CAS Key Laboratory of Strongly-Coupled Quantum Matter Physics, and Department of Physics, University of Science and Technology of China, Hefei, Anhui 230026, China}
\author{Yulei Han$^\dag$}
\affiliation{ICQD, Hefei National Laboratory for Physical Sciences at Microscale, CAS Key Laboratory of Strongly-Coupled Quantum Matter Physics, and Department of Physics, University of Science and Technology of China, Hefei, Anhui 230026, China}
\author{Yafei Ren$^\dag$}
\affiliation{Department of Physics, The University of Texas at Austin, Austin, Texas 78712, USA}
\author{Qian Niu}
\affiliation{Department of Physics, The University of Texas at Austin, Austin, Texas 78712, USA}
\author{Zhenhua Qiao}
\email[Correspondence author:~~]{qiao@ustc.edu.cn}
\affiliation{ICQD, Hefei National Laboratory for Physical Sciences at Microscale, CAS Key Laboratory of Strongly-Coupled Quantum Matter Physics, and Department of Physics, University of Science and Technology of China, Hefei, Anhui 230026, China}	
\date{\today{}}
	
\begin{abstract}	
We identify a valley-polarized Chern insulator in the van der Waals heterostructure,  Pt$_{2}$HgSe$_{3}$/CrI$_3$, for potential applications with interplay between electric, magnetic, optical, and mechanical effects.
The interlayer proximity  magnetic coupling nearly closes the band gap of Pt$_{2}$HgSe$_{3}$ and the strong intra-layer spin-orbit coupling further lifts the valley degeneracy by over 100 meV leading to positive and negative band gaps at opposite valleys. In the valley with negative gap, the interfacial Rashba spin-orbit coupling opens a topological band gap of 17.8 meV, which is enlarged to 30.8 meV by adding an $h$-BN layer.
We find large orbital magnetization in Pt$_{2}$HgSe$_{3}$ layer that is much larger than spin, which can induce measurable optical Kerr effect.
The valley polarization and Chern number are coupled to the magnetic order of the nearest neighboring CrI$_3$ layer, which is switchable by electric, magnetic, and mechanical means in experiments.
The presence of $h$-BN protects the topological phase allowing the construction of superlattices with valley, spin, and layer degrees of freedoms.
\end{abstract}
	
\maketitle
	
\textit{Introduction.---}
The inequivalent electronic band extrema of graphene at K and K$'$ inspired the development of valleytronics~\cite{Valleytronics_07}, which encodes information in the valley binary and focuses on its generation, manipulation, and detection~\cite{Rev_Valleytronics_16, Rev_Valleytronics_18}. With breaking inversion symmetry, the valley contrasting orbital magnetic moment and valley Hall effect have been demonstrated
and materialized in transition metal dichalcogenides (TMD) monolayers~\cite{ValleyContrasting_07, ValleyContrasting_TMD_12}. In these materials, the generation and manipulation of valley polarization  have been realized by optical or magnetic fields~\cite{TMD_band_15, TMD_Optical_Mak_12, TMD_Optical_Cao_12, TMD_Optical_Zeng_12, TMD_Optical_Jones_12, TMD_Optical_Mak_14, TMD_Magnetic_Kim_14, TMD_Magnetic_Srivastava_15, TMD_Magnetic_Rostami_15, TMD_Magnetic_Aivazian_15, TMD_Magnetic_Stier_15, TMD_Magnetic_MacNeill_15, TMD_Magnetic_Li_15,ValleyLayer_SYYang_20}. Despite such successes in experiments, these methods are difficult to be scaled, which limits the potential future application. Electrical control of valley degree of freedom, which is scalable and more compatible with the current semiconductor technology, is thus still highly desired.
	
The electrical generation of valley polarization can be achieved, when the valley degeneracy is lifted by breaking the time reversal symmetry~\cite{TMD_Magnetic_Kim_14, TMD_Magnetic_Srivastava_15, TMD_Magnetic_Rostami_15, TMD_Magnetic_Aivazian_15, TMD_Magnetic_Stier_15, TMD_Magnetic_MacNeill_15, TMD_Magnetic_Li_15} that can be achieved by applying magnetic field or proximity coupling to magnetic substrates~\cite{TMD_MagSub_MoTe2EuO_15, TMD_MagSub_MoTe2EuO_16, TMD_MagSub_MoS2EuS_17, TMD_MagSub_WSe2EuS_Exp_17, TMD_MagSub_WS2EuS_Exp_19, TMD_MagSub_WS2MnO_18, TMD_MagSub_TMDMnO_18, TMD_MagSub_WS2MnO2_18, TMD_MagSub_WTe2YMnO3_17, TMD_MagSub_MoS2CoO_18, TMD_MagSub_MoTe2Fe3O4_16, TMD_vdWMag_WSe2CrI3_Exp_18, TMD_vdWMag_MoWSe2CrI3_19, TMD_vdWMag_WS2hVN_19}. By using bulk magnetic substrate, valley splitting in TMD layers is observed, but the magnitude is limited to several meV~\cite{TMD_MagSub_WSe2EuS_Exp_17, TMD_MagSub_WS2EuS_Exp_19}. Moreover, to manipulate the valley polarization one needs to switch the magnetization of the bulk substrate, which is however very challenging to be achieved by electrical means.
%
In contrast, recent development on two-dimensional van der Waals (vdW) magnets~\cite{Rev_vdWLayers_17, CrI3_ElectricalSwitch_18, Rev_Valleytronics_Control_19, Rev_Proximity_19, Rev_2D_FM_20, Rev_2D_FM_proximity_20} opens up such possibility~\cite{TMD_vdWMag_WSe2CrI3_Exp_18, TMD_vdWMag_MoWSe2CrI3_19, TMD_vdWMag_WS2hVN_19, TMD_vdWMag_MoTe2WSe2CrX3_20, Rev_Proximity_19, Rev_2D_FM_20, Rev_2D_FM_proximity_20}.
Switching of magnetization of CrI$_3$ few layers have been achieved in experiments by applying electric or magnetic fields or external strain~\cite{CrI3_ElectricalSwitch_18, Rev_Valleytronics_Control_19}, which allows the construction of valley-field effect transistor and other devices with interplay between electric, magnetic, optical, and mechanical effects.

A proper material candidate showing large valley polarization induced by CrI$_3$ is however still lacking despite several attempts.
In TMD/CrI$_3$ heterostructure, the valley splitting is limited to few meV due to the band alignment faulty~\cite{TMD_vdWMag_WSe2CrI3_Exp_18, TMD_vdWMag_MoWSe2CrI3_19, TMD_vdWMag_MoTe2WSe2CrX3_20}.
Besides, the energy bands of CrI$_3$ lie inside the band gap of TMD materials. This alters the intrinsic band structure of TMD layers and can even close its band gap~\cite{TMD_vdWMag_MoTe2WSe2CrX3_20}. In stanene/CrI$_3$ heterostructure~\cite{CrI3_Stanene_19}, though the valley splitting energy bands are realized, they show strong overlapping with the substrate's bands. 

In this Letter, we propose an ideal material system, Pt$_{2}$HgSe$_{3}$/CrI$_3$, showing large valley splitting and nonzero Chern number. In the absence of spin-orbit coupling (SOC), the interlayer magnetic proximity effect nearly close the band gap of Pt$_{2}$HgSe$_{3}$. The strong intrlayer SOC induces energy gaps with opposite signs in K and K$'$ valleys leading to large valley splitting over 100 meV. The interfacial Rashba SOC further opens a gap of 17.8 meV with nonzero Chern number in the negatively gapped valley. In Pt$_{2}$HgSe$_{3}$ layer, we identify a large orbital magnetic moment that is larger than that from spin, which can induce orbital-magnetization based optical Kerr effect. 
We find that the valley polarization and Chern number are coupled to the magnetism of the nearest neighboring CrI$_3$ and are robust against the increasing of the CrI$_3$ thickness.
In the presence of an $h$-BN layer, we find that the topological band gap can be enlarged to 30.8 meV whereas the valley splitting remains large. The protection provided by $h$-BN layer further allows the construction of superlattices with valley, spin and layer degree of freedom.
	
\begin{figure}
  \includegraphics[width=8cm,angle=0]{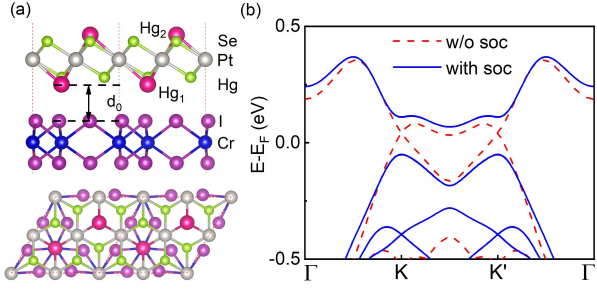}
  \caption{(a) The structure of Pt$_2$HgSe$_3$/CrI$_3$ heterostructure. The pink, silvery, green, blue and purple balls are for Hg, Pt, Se, Cr and I atoms, respectively. $d_0$ represents the optimized vdW gap and the two buckled Hg sites are respectively denoted as Hg$_1$, Hg$_2$. (b) Band structure of monolayer Pt$_2$HgSe$_3$ without and with SOC, respectively.}
  \label{Fig1}
\end{figure}

\textit{Atomic Structures.---} We focus on the Pt$_2$HgSe$_3$-based heterostructure as displayed in Fig.~\ref{Fig1}(a). The bulk Pt$_{2}$HgSe$_{3}$ is experimentally demonstrated as a dual-topological semimetal~\cite{PHS exp1}, which can be exfoliated down to few layers that are expected to be stable under ambient condition~\cite{PHS exp2}. The Hg atoms, which contribute dominantly to the energy bands near the Fermi energy, form a buckled honeycomb lattice. The electronic structure of a free-standing Pt$_{2}$HgSe$_{3}$ monolayer is plotted in Fig.~\ref{Fig1}(b). In the absence of spin-orbit coupling, linear dispersion exists around K and K$'$ points by dashed lines. When the spin-orbit coupling is further included, a large band gap opens as plotted by solid lines, which indicates that Pt$_{2}$HgSe$_{3}$ can be considered as a Kane-Mele type topological insulator~\cite{PHS1,PHS2,PHS3}.
	
By placing Pt$_{2}$HgSe$_{3}$ on CrI$_3$, the magnetism can be induced from the proximity effect, which depends strongly on the atomic wavefunction overlap. Thus, the two sublattices of Hg atoms are expected to experience different exchange fields. We denote the sublattice of Hg near CrI$_3$ as Hg$_1$ whereas the other sublattice is Hg$_2$. In our calculation, we adopt a $1\times 1$ Pt$_2$HgSe$_3$/CrI$_3$ unit cell with a lattice mismatch of $4.5\%$.

\textit{Effective Model of Valley Splitting.---} Before demonstrating detailed first-principles calculation results, the analysis based on an effective model helps to illustrate the underlying physics clearly. The electronic structure of Pt$_{2}$HgSe$_{3}$ can be described by the Kane-Mele model of $H_0 + H_{\rm ISO}$, where $H_0 = \hbar v_{\mathrm{F}}(\tau_z \sigma_x k_x + \sigma_y k_y)s_0$ describes its Dirac dispersion in the absence of spin-orbit coupling with Fermi velocity $v_{\rm F}$ and $H_{\rm ISO} = \lambda \tau_z \sigma_z s_z$ is the intrinsic spin-orbit coupling with $\lambda = 81.2~\mathrm{meV}$. $\bm{s}$, $\bm{\sigma}$, and $\bm{\tau}$ are Pauli matrices for spin, sublattice and valley, respectively.
	
\begin{figure}
  \includegraphics[width=8.5cm,angle=0]{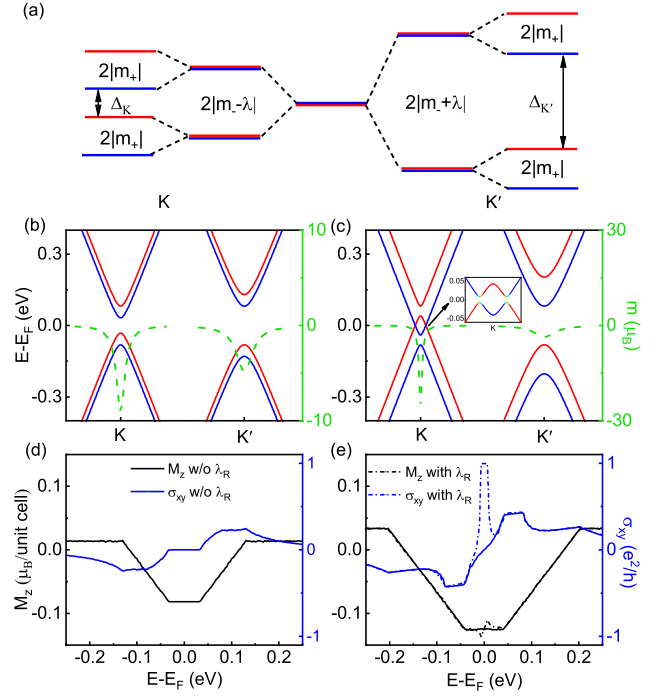}
  \caption{(a) The influence of symmetric and anti-symmetric exchange field on the band gaps in K and K$'$ valleys. (b)-(c) The band structure and orbital magnetic moment around K and K$'$ valleys for (b) $m_1=0.6\lambda$ and (c) $m_1=1.5\lambda$, respectively. The inset of (c) shows the band near Fermi energy with inclusion of small Rashba spin-orbit coupling. The red (blue) color denotes spin up (down) state. (d)-(e) The corresponding orbital magnetization in an unit cell and anomalous Hall conductance. The hollow (solid) square represent the system without (with) Rashba spin-orbit coupling.}
  \label{Fig2}
\end{figure}
	
The presence of magnetic substrate breaks both inversion ($\mathcal{P}$) and time-reversal ($\mathcal{T}$) symmetries and leads to inequivalent exchange fields $m_{1,2}$ on Hg$_1$ and Hg$_2$, which can be described by $H_{\rm EX}=m_+ \sigma_0 s_z   + m_- \sigma_z s_z$ where $m_{\pm}=(m_1\pm m_2)/2$ stand for symmetric and anti-symmetric parts of the inequivalent exchange fields. The anti-symmetric part together with $H_{\rm ISO}$ give rise to a mass term $(m_- + \lambda \tau_z)\sigma_z s_z$ where $s_z$ indicates spin-contrast sign of mass leading to, in contrast to TMD, a vanishing valley Chern number. The presence of $\tau_z$ leads to valley-dependent band gaps as illustrated in Fig.~\ref{Fig2}(a). In this case, the spin up and down bands are still degenerate since the $\mathcal{P}\mathcal{T}$ symmetry remains, which also guarantees a vanishing magnetization and anomalous Hall response. When the symmetric part $m_+$ is further included, $\mathcal{P}\mathcal{T}$ symmetry is broken that lifts the spin degeneracy and valley-dependent band gaps. Nonzero orbital magnetization and anomalous Hall responses can also appear.
	
By fitting to the first-principles result, we find that $|m_1| \gg |m_2|$ since the Hg$_2$ atoms are further away from the magnetic substrate. We thus set $m_2=0$ for simplicity. With a moderate exchange field $|m_1|< \lambda$, we plot the energy bands in Fig.~\ref{Fig2}(b) in solid lines where one can find a spin-valley splitting. These valley-splitting energy bands exhibit large orbital magnetic moment as plotted in green dashed lines where the magnetic moment of the valence band maximum can reach about 9.0 $\mu_\mathrm{B}$ (4.9 $\mu_\mathrm{B}$) around K (K$'$) point for $m_1=0.6\lambda$. When the Fermi energy lies inside the band gap, finite orbital magnetization appears in the order of $0.1~\mu_B$ per unit cell as shown in Fig.~\ref{Fig2}(d) that is larger than the spin magnetization of Hg atoms. The presence of orbital magnetization suggests the presence of anomalous Hall effect, Kerr effect etc~\cite{BerryReview_2010,OrbitalMag_2007,KerrEff_15,OrbitalMag_2015,OrbitalMag_2019}.
	
The valley splitting gradually increases as the inequivalent exchange field increases as plotted in Fig.~\ref{Fig2}(c). Meanwhile, the larger $m_1$ makes the effective mass smaller in valley $K$, resulting in a larger orbital magnetic moment that is about 25.2 $\mu_\mathrm{B}$ for $m_1 = 1.5\lambda$. The corresponding orbital magnetization and anomalous Hall effect are also stronger as illustrated in Fig.~\ref{Fig2}(e). More importantly, when $|m_+| > |m_- - \lambda|$, the band inversion appears between two bands with opposite spins. In this case, the interfacial Rashba spin-orbit coupling $H_{\mathrm{R}} = \lambda_{\mathrm{R}} (\tau_z \sigma_x s_y - \sigma_y s_x)$ plays an essential role, which reopens a band gap in $K$ valley as shown in the inset, which harbours the quantum anomalous Hall effect (QAHE) with a Chern number of $\mathcal{C}=1$. Such topological property is demonstrated by the quantized Hall conductance and linear dependence of magnetization on chemical potential inside the band gap.
	
\begin{figure}
  \includegraphics[width=8.5cm,angle=0]{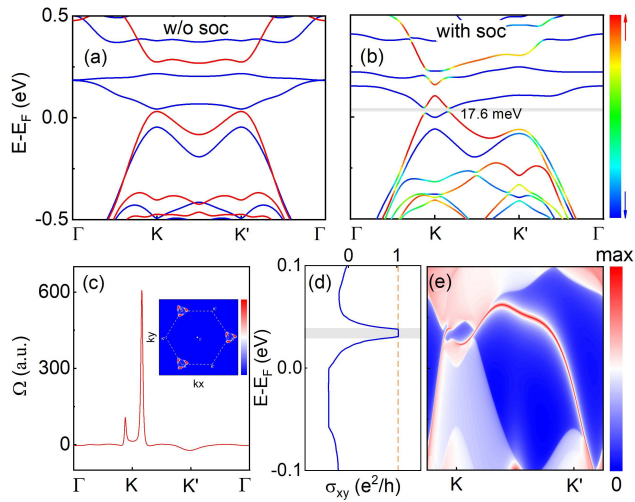}
  \caption{(a)-(b) Band structure of Pt$_2$HgSe$_3$/CrI$_3$ without (a) and with (b) SOC. In (a), the red (blue) represents spin up (down) state. In (b), $\Delta_{K}=-105.5~$meV and $\Delta_{K'}=167.1~$meV with valley splitting $\Delta=(\Delta_{K'}-\Delta_{K})/2=136.3~$meV. The gap with negative sign indicates the presence of band inversion.
  (c) The Berry curvature along the high symmetry lines. The inset shows the Berry curvature in the first Brillouin zone. (d) The Fermi energy dependence of the anomalous Hall conductivity. (e) The local density of states of a semi-infinite zigzag ribbon of Pt$_{2}$HgSe$_{3}$/CrI$_{3}$ heterostructure.}\label{Fig3}
\end{figure}

\textit{Fully Valley-polarized QAHE in  Pt$_2$HgSe$_3$/CrI$_3$.---} The first-principles calculations agree well with our theoretical analysis. The valley-polarized QAHE can really be formed in Pt$_{2}$HgSe$_{3}$/CrI$_3$ heterostructure as displayed in Figs.~\ref{Fig3}(a) and \ref{Fig3}(b), where we plot the electronic structures without and with spin-orbit coupling, separately. In the absence of spin-orbit coupling, the magnetic proximity effect leads to a large spin splitting, while the valley degeneracy is preserved. The nearly flat band above the Fermi level arises from the hybridization with CrI$_3$ bands, but does not affect the physics described above~\cite{SM}. When the spin-orbit coupling is further included, a strong valley splitting of 136.3 meV is observed as shown in Fig.~\ref{Fig2}(b), where a band inversion appears at K valley with a band gap of 17.6 meV. The spin projection in color agrees with our above model, indicating that the band gap is opened by the Rashba spin-orbit coupling and is topological nontrivial.
	
To confirm the band topology, we calculate the Berry curvature by using maximally-localized Wannier functions~\cite{wannier90}. We find that the Berry curvature shows sharp positive peaks in K valley, whereas becomes vanishingly small at K$'$ valley as shown in Fig.~\ref{Fig3}(c).
By integrating the Berry curvature around K/K$'$ valleys, we find that the valley-dependent Chern numbers are $\mathcal{C}_K=1$ and $\mathcal{C}_{K'}=0$. Thus the corresponding total Chern number is $\mathcal{C}=\mathcal{C}_K+\mathcal{C}_{K'}=1$ and the valley Chern number is $\mathcal{C}_V=\mathcal{C}_K - \mathcal{C}_{K'}=1$, indicating the presence of a fully valley-polarized QAHE. The energy dependence of the anomalous Hall conductance is plotted in Fig.~\ref{Fig3}(d), which shows a quantized value of $e^2/h$ in the energy gap. Such a topological phase is further confirmed by studying the topological edge states~\cite{green,wanniertools}. As shown in Fig.~\ref{Fig2}(e), we find two gapless edge modes with positive group velocity at K valley connecting the valence and conduction bands, whereas one edge state in K$'$ valley with opposite velocity. The number difference of the left and right moving edge states in different valleys agrees with the Chern number as well as the valley Chern number.
	
Besides the band topology, the first principles calculations can provide further details about the magnetism in the heterostructure. The proximity effect leads to different magnetic moments at two Hg atoms, i.e., the local magnetic moment is 0.015 $\mathrm{\mu_B}$ in ``Hg$_1$" that is parallel to the CrI$_3$, whereas a magnetic moment of -0.017 $\mathrm{\mu_B}$ is induced in ``Hg$_2$" with opposite sign. It is noteworthy that the net spin magnetic moment is three orders of magnitude smaller than that from orbital contribution, indicating that the large orbital magnetic moment plays a crucial role in the Pt$_2$HgSe$_3$-based system. In addition, the local magnetic moment of Cr is also increased by 0.5 $\mathrm{\mu_B}$ comparing to that in pristine CrI$_3$ monolayer (3.0 $\mathrm{\mu_B / Cr}$)~\cite{BiSe CrI3}. As a result, the Heisenberg exchange constant increases to $J_0=-5.23$ meV that corresponds to a Curie temperature of 91 K~\cite{SM}. Therefore, the magnetism can be enhanced in the heterostructure, benefiting the realization of high-temperature valley-polarized QAHE.
	
\begin{figure}			
	\includegraphics[width=8.5cm,angle=0]{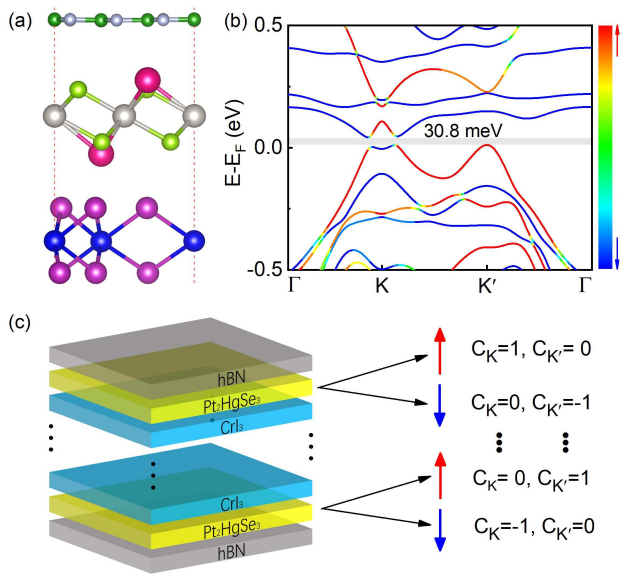}
	\caption{(a) Lattice and (b) electronic structures of $h$-BN/Pt$_{2}$HgSe$_{3}$/CrI$_3$ with SOC. Here, $\Delta_{K}=-113.4~$meV and $\Delta_{K'}=137.4~$meV with valley splitting $\Delta=125.4~$meV. The gap with negative sign indicates the presence of band inversion. (c) vdW superlattice consisting of $h$-BN/Pt$_{2}$HgSe$_{3}$/CrI$_3$ multilayer as building blocks to realize abundant topological phases coupled spin, valley and layer degree of freedoms with Chern numbers.}\label{Fig4}
\end{figure}

\textit{$h$-BN/Pt$_{2}$HgSe$_{3}$/CrI$_3$ as a Building Block.---} To study the scalability of the heterostructure, we investigate the influence of $h$-BN on the topological phase since $h$-BN is widely used in experiments as a dielectric layer, which is atomically flat, stable in air and high temperature, and can protect the vdW layers from contamination and degradation~\cite{hBN1,hBN2}. By considering the trilayer system as illustrated in Fig.~\ref{Fig4}(a), we plot the energy bands with spin-orbit coupling in Fig.~\ref{Fig4}(b), where the large valley splitting remains and the topological band gap is nearly doubled to $30.8~$meV, larger than the one without $h$-BN. As the electrical control of the magnetism of CrI$_3$ is experimentally found in its bilayers, we also studied the Pt$_{2}$HgSe$_{3}$/CrI$_3$ bilayer structure and find similar results~\cite{SM}, which greatly improves the application potentials of the vdW layers studied here.
	
The enhanced valley-polarized QAHE by encapsulated $h$-BN layer and its robustness with increasing CrI$_3$ thickness not only make the QAHE achievable at higher temperature, but also make the $h$-BN/Pt$_{2}$HgSe$_{3}$/CrI$_3$ a perfect building block to realize multi-functional electronics. By reversing the stacking order from $h$-BN/Pt$_{2}$HgSe$_{3}$/CrI$_{3}$ to CrI$_{3}$/Pt$_{2}$HgSe$_{3}$/$h$-BN, one can change the valley polarization without changing the Chern number. Moreover, with the stacking order fixed, the valley polarization and Chern number can be simultaneously reversed by reversing the magnetization. Therefore, by changing the stacking orders and magnetization directions, one can control the spin, valley, and layer degree of freedoms with Chern number in a superlattice structure as illustrated in Fig.~\ref{Fig4}(c). Rich structures with different functionalities in the spin- and valleytronics are expected.
	
\textit{Summary.---} We propose the vdW heterostructure Pt$_{2}$HgSe$_{3}$/CrI$_3$ as an ideal material system to realize topological valleytronics with interplay between electric, magnetic, optical, and mechanical effects. The imbalanced proximity exchanged field and the intrinsic strong SOC of Pt$_{2}$HgSe$_{3}$ lead to a large valley splitting over 100 meV. In the valley with band inversion, the interfacial Rashba SOC leads to a Chern insulator with a band gap of 17.8 meV, which can be enlarged to 30.8 meV in the presence of $h$-BN. The valley-encoded dissipationless edge states benefits the realization of low-energy consumption topological devices. Large orbital magnetization and negligible spin magnetization are identified in Pt$_{2}$HgSe$_{3}$, which can lead to orbital effect dominated optical Kerr effect and anomalous Hall effect. This not only benefits the electrical and optical detection of the valley index but also provides the opportunity to switch the magnetism and valley polarization via electric current at finite doping.

The valley polarization and Chern number are coupled to the magnetism of the nearest neighboring CrI$_3$ and are robust against the increasing of the CrI$_3$ thickness.
The experimentally demonstrated switching of the magnetism in CrI$_3$ few layers by electric, magnetic, or mechanical means opens up the possibility to realize valleytronic devices controlled by these methods.
Particularly, the electrical switching of the magnetism lays the foundation of its application in electrically controllable valleytronics, e.g., the valley-field effect transistor, which can combine the advantage of nontrivial band topology and dissipationless kink states at the domain walls.
Moreover, the robustness of the topological phase in the presence of an $h$-BN layer opens up the possibility to construct superlattices with valley, spin and layer degrees of freedom.


	

Besides CrI$_3$, we find that other magnetic insulators, e.g., MnI$_2$, can also be used as substrate materials to realize the valley-polarized QAHE in Pt$_{2}$HgSe$_{3}$-based vdW heterostructures~\cite{SM}. Considering the continuously growing family of vdW layers with different functionalities, e.g., magnetism, large-gap dielectric materials, large-gap $Z_2$ topological insulator in jacutingaite-family~\cite{Jacutingaite}, our results suggest a way to search material candidates with novel physics and application potentials in topological valleytronics.
	
\begin{acknowledgments}
\textit{Acknowledgments.---} This work was financially supported by the National Key Research and Development Program (2017YFB0405703), the National Natural Science Foundation of China (11974327), China Postdoctoral Science Foundation (2020M681998), Anhui Initiative in Quantum Information Technologies, and the Fundamental Research Funds for the Central Universities. We are grateful to AMHPC and Supercomputing Center of USTC for providing the high-performance computing resources.
YFR was supported by the Welch Foundation
(F-1255), and QN by DOE (DE-FG03-02ER45958, Division of Materials Science and Engineering) on general
theoretical considerations.
\end{acknowledgments}

$^{\ddag}$  Z. Liu, Y. Han and Y. Ren contributed equally to this work.

\end{document}